# Modification of structural and mechanical properties of diamond with MeV ion implantation

F.Bosia[1], N.Argiolas[2], M.Bazzan[2], P.Olivero[1], F.Picollo[1], A.Sordini[3], M.Vannoni[3], H.Wang[1], E.Vittone[1].

[1]Dept. Experimental Physics- NIS Centre of Excellence, Università di Torino, INFN Torino, Italy

[2] Dept. Physics, Università di Padova, Italy

[3]INO-CNR Firenze, Italy

**Abstract**

We present experimental results and numerical simulations to investigate the modification of structural-mechanical properties of ion-implanted single-crystal diamond. A phenomenological model is used to derive an analytical expression for the variation of mass density and elastic properties as a function of damage density in the crystal. These relations are applied together with SRIM Monte Carlo simulations to set up Finite Element simulations for the determination of internal strains and surface deformation of MeV-ion-implanted diamond samples. The results are validated through comparison with high resolution X-ray diffraction and white-light interferometric profilometry experiments. The former are carried out on 180 keV B implanted diamond samples, to determine the induced structural variation, in terms of lattice spacing and disorder, whilst the latter are performed on 1.8 MeV He implanted diamond samples to measure surface swelling. The effect of thermal processing on the evolution of the structural-mechanical properties of damaged diamond is also evaluated by performing the same profilometric measurements after annealing at 1000 °C, and modeling the obtained trends with a suitably modified analytical model. The results allow the development of a coherent model describing the effects of MeV-ion-induced damage on the structural-mechanical properties of single-crystal diamond. In particular, we suggest a more reliable method to determine the so-called diamond "graphitization threshold" for the considered implantation type.

# 1. Introduction

Considerable effort has been devoted in recent years to the application of high energy (MeV) ion implantation in the fabrication and functionalization of diamond, in particular with the aim of developing a series of micro-devices, ranging from bio-sensors and detectors to micro-electromechanical systems (MEMS) and optical devices [1-3]. This can be achieved by exploiting the strongly non-uniform damage depth profile of MeV ions and creating specific regions where the diamond lattice structure is critically damaged. After annealing, graphitization of the buried layer is achieved whilst the remaining surrounding material is restored to pristine diamond, so that spatially well-defined structures can be created by subsequently etching away the graphitized regions [4] or advantage can be taken from the conductive properties of the graphitized regions [*il lavoro nostro e quello di Sharkov tra le citazioni che ho suggerito*].

In order to control the spatial extension of the graphitized layer with some accuracy, it is necessary to acquire in depth knowledge of how the diamond lattice structure is modified as a function of implanted ion type, energy, fluence, implantation temperature, annealing temperature, local stress, etc. In particular, a critical damage level $N_C$ has been identified in the literature, above which diamond is subject to permanent amorphization and subsequent graphitization upon thermal annealing [5], although some uncertainty remains on the value of $N_C$ and its dependence on implantation parameters (e.g. depth and/or local strain, self-annealing, etc).

A well-known effect of ion implantation in diamond is surface swelling in correspondence with the irradiated regions, due to the internal density variation in the damaged crystal, which causes a constrained volume expansion [6]. A detailed analysis of this effect can provide additional information on the structural modifications occurring in ion-implanted diamond. This type of study was the object of investigation in [7], where a phenomenological model accounting for saturation in vacancy density was developed, and Finite-Element (FEM) Simulations were performed to compare numerical results with experimental surface swelling measurements. In this paper, we extend the numerical procedure to model annealing effects in ion-implanted specimens, introducing the concept of a graphitization threshold in a rigorous analysis. To validate the model, the results of simulations were compared with surface swelling profilometry measurements and with high resolution X-ray diffraction (HR-XRD) analyses.

The paper is structured as follows: in Section 2, the model for structural variation of damaged diamond is reviewed and extended to annealed specimens; in Section 3, Experimental and numerical techniques are presented; in Section 4, HD-XRD and profilometric measurements are presented and discussed, and the numerical fitting of experimental swelling results is carried out.

## 2. Model

We adopt the model presented in [7], which accounts for saturation in the creation of vacancies in the damaged diamond crystal lattice for increasing fluencies in ion implantation. The vacancy density of the damaged diamond as a function of substrate depth $z$ and implantation fluence $F$ can be expressed as:

$$\rho_V(F,z) = \alpha \left(1 - e^{-\frac{\lambda(z)F}{\alpha}}\right) \quad (1)$$

where the constant $\alpha$ represents the vacancy density saturation value, which depends on the implantation type, and $\lambda$ is the linear vacancy depth profile calculated using the SRIM 2008.04 code [8]. The corresponding mass density $\rho$ of the damaged diamond is:

$$\rho(F,z) = \rho_d - (\rho_d - \rho_{aC})\left(1 - e^{-\frac{\lambda(z)F}{\alpha}}\right) \qquad (2)$$

where $\rho_d =3.52$ g cm$^{-3}$ is the diamond density and $\rho_{aC} =1.56$ g cm$^{-3}$ is the amorphous carbon density []. Similarly for the substrate Young's modulus:

$$E(F,z) = E_d - (E_d - E_{aC})\left(1 - e^{-\frac{\lambda(z)F}{\alpha}}\right) \qquad (3)$$

with $E_d =1220$ GPa and $E_{aC} =10$ GPa, and Poisson's ratio:

$$v(F,z) = v_d - (v_d - v_{aC})\left(1 - e^{-\frac{\lambda(z)F}{\alpha}}\right) \qquad (4)$$

with $v_d =0.2$ and $v_{aC} =0.45$ [9].

When damaged diamond is annealed between 1000°-1200°C, the crystal lattice is almost completely restored to that of pristine diamond if the vacancy density is below the so-called graphitization threshold $N_C$, whilst there is a transition to the structure and properties of graphite for material portions above $N_C$. Therefore, to determine the density depth profile for a given implanted substrate, one must compare the corresponding vacancy density profile with the graphitization threshold $N_C$. An example is shown in

Fig. 1 for 1.8 MeV He$^+$ implantations at fluences of $F_1=1.4 \cdot 10^{16}$ ions cm$^{-2}$ and $F_2=10.5 \cdot 10^{16}$ ions cm$^{-2}$. For the implantation fluence $F_1$, the damage density curve $\rho_V(z)$ lies entirely below $N_C$ ("below threshold" in Fig. 1), and the structural/mechanical properties before annealing are given by Eqs. 2-4, whilst after annealing they are restored to those of diamond. For the implantation fluence $F_2$, the damage density curve has a peaked damage density lying above $N_C$ ("above threshold" in Fig. 1), and the structural/mechanical properties before annealing are again given by Eqs. 2-4, whilst after annealing they are converted to those of graphite (i.e. $\rho_g = 1.80$ g cm$^{-3}$, $E_g = 20$ GPa, $v_g = 0.2$ g cm$^{-3}$) for $\rho_V(z) > N_C$ and restored to those of diamond for $\rho_V(z) < N_C$.

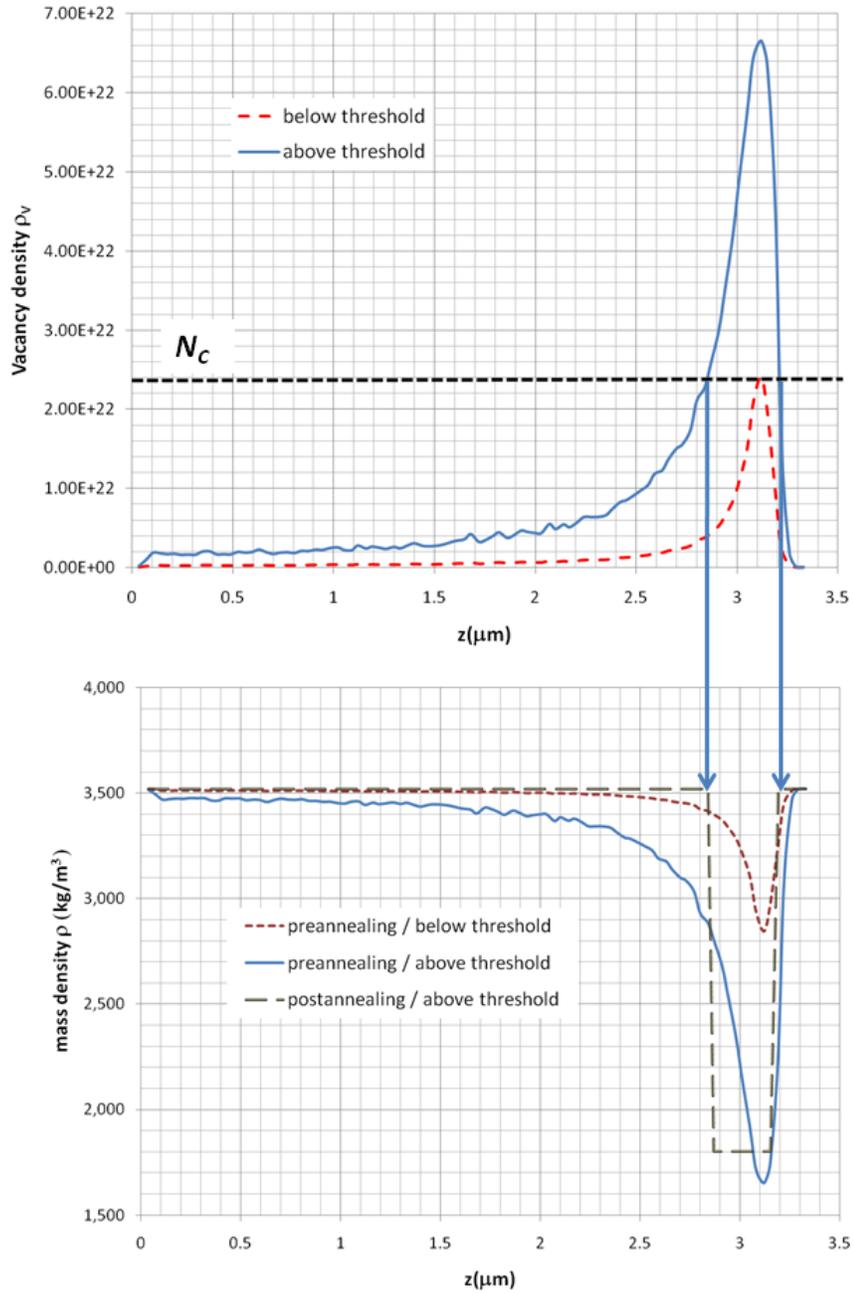

Fig. 1: a) Vacancy density vs. specimen depth for diamond implanted with 1.8 MeV He ions: two examples are shown for $F_1=1.4 \cdot 10^{16}$ ions cm$^{-2}$ and $F_2=10.5 \cdot 10^{16}$ ions cm$^{-2}$, below and above the graphitization threshold, respectively; b) corresponding mass density before and after annealing at 1000 °C.

## 3. Experimental and numerical techniques

### 3.1. Samples and ion implantation

Ion implantation was performed on HPHT (produced by Sumitomo, type Ib, 100 crystal orientation) and CVD (produced by ElementSix, type IIa, 100 crystal orientation) samples, 3×3×1.5 mm$^3$ in size, with two optically polished opposite large faces. The two sample types, HPHT and CVD, can be reasonably expected to display the same high-quality single-crystal mechanical properties.

In order to use the HD-XRD technique, diamond samples with relatively shallow ion implantations are required, as the probing depth extends to about 500 nm. Therefore, the first (CVD) sample was implanted with 180 keV B ions, at the Olivetti I-Jet facilities in Arnad (Aosta, Italy). The whole upper surface of the sample was irradiated uniformly with a fluence of $10^{13}$ ions cm$^{-2}$.

The second (HPHT) sample was irradiated with 1.8 MeV He ions at the ion microbeam line of the INFN Legnaro National Laboratories (Padova, Italy). Typically, ~125×125 $\mu$m$^2$ square areas were implanted by raster scanning an ion beam with size of 20-30 $\mu$m. The implantation fluences, ranging from $1 \cdot 10^{16}$ ions cm$^{-2}$ to $2 \cdot 10^{17}$ ions cm$^{-2}$, were controlled in real time by monitoring the X-ray yield from a thin metal layer evaporated on the sample surface. The implantations were performed at room temperature, with ion currents of ~1 nA.

### 3.2. HD-XRD measurements

Boron-implanted samples were investigated with the HR-XRD technique at the department of Physics of the University of Padova. The instrument used is a Philips

MRD diffractometer. The source is an X-ray tube with copper anode, equipped with a parabolic mirror and a Ge (2 2 0) Bartels monochromator with four rebounds. The resulting primary beam has a divergence angle of 0.0039° and a spectral purity of $\Delta\lambda/\lambda = 10^{-5}$ at a wavelength $\lambda = 1.54056$ Å. The beam impinges on the sample at an angle $\omega$ measured by a high precision goniometer, on which the specimen is mounted. The scattered radiation from the sample is measured as a function of the incidence angle $\omega$ and the scattering angle $2\theta$ by a Xe proportional detector mounted on a second independent goniometer, coaxial to the first. To improve the resolution of the angular measurement, the detector is equipped with a Ge (2 2 0) analyzer with three rebounds that guarantees acceptance angle of 0.0039°. The system is maintained in a measuring chamber at a constant temperature of (25.0 ± 0.1) °C *[reference!]*.

The unimplanted and implanted HPHT and CVD samples are measured under the same conditions *[cosa vuol dire?]* near the (0 0 4) reflection. The measurement consists in a $\omega$-$2\theta$ scan so that the incidence and collection angles are varied in a synchronized manner.

### 3.3. Profilometry measurements

Surface swelling data were acquired at the Interferometry laboratories of the Istituto Nazionale di Ottica (INO), Firenze, with a Zygo NewView 6000 system, which exploits white light interferometry to provide detailed, non contact measurements of 3-D profiles. A vertical resolution of 0.1 nm was achieved over a lateral range up to 150 µm, while lateral resolution varied from 4.6 µm up to 0.6 µm, depending on the objective.

### 3.4. FEM simulations

Simulations were carried out using "Structural mechanics" module COMSOL Multiphysics, as explained in [7]. Specimen geometry is reproduced and meshed in 2-D simulations. The analytical expressions for density and mechanical parameters of the substrate given by Eq. 2-4 are used to model specimens before annealing, together with the damage profile $\lambda(z)$ resulting from SRIM simulations for 180 keV B and 1.8 MeV He, with a displacement energy of a carbon atom in diamond of 50 eV []. The saturation value for vacancy density $\alpha$ for the two implantation types are extrapolated by best fitting procedures to be $\alpha_{180\ keV\ B} = 8 \cdot 10^{22}$ ions cm$^{-2}$ and $\alpha_{1.8\ MeV\ He} = 7 \cdot 10^{22}$ ions cm$^{-2}$. For annealed specimens, the step-like density and mechanical parameter depth variation illustrated in Fig.1 is used. For both implantations types, surface deformation and internal strains are calculated as a function of the irradiation fluence.

## 4. Results and discussion

### 4.1. HD-XRD measurements

In Figure 2a we report the $\omega$-$2\theta$ rocking curve for unimplanted diamond substrates acquired in proximity of the (0 0 4) reciprocal lattice point. From the comparison between the two substrates and a reference calculation, it can be seen that both the substrates are characterized by a large FWHM value. As this is several times larger than the theoretical peak width calculated on the basis of the dynamical theory of X-ray diffraction corrected for experimental broadening, it indicates the presence of some kind of structural disorder in the substrate prior to implantation. A series of $\omega$–scans as a function of the beam position on the sample reveals that this is due to bending of the lattice planes in the sample [10]: the maximum curvature is found near the sample edges and can be estimated to have a radius of about 32 m. Thus, the substrates are affected prior to the implantation by the presence of slowly varying structural disorder, probably caused by the presence of residual thermal stresses which deform the sample, and result in a wider Bragg peak in HD-XRD measurements.

The reciprocal space map for the implanted samples with B at 180 keV (not shown) provides evidence for implantation–induced deformation, with no relaxation occurring at the interface between the implanted region and the underlying substrate. The deformation is positive, i.e. the implantation provokes lattice expansion, as expected. The $\omega$-$2\theta$ scan acquired along the symmetrical (0 0 4) lattice point for the substrate implanted with a fluence of $10^{13}$ ions cm$^{-2}$ is reported in Figure 2b. The shoulder appearing in the Bragg peak proves the presence of lattice expansion *[questo bisognerebbe giustificarlo, ad es con referenza]*. To determine the value of implantation-induced lattice expansion, these experimental results are compared to simulated profiles obtained using the dynamical theory of X-rays diffraction, approximating the deformation profile as a series of lamellae characterized by a fixed thickness and a homogeneous deformation along the vertical direction. Due to the above–mentioned curvature of the substrate, the experimental rocking curves must be thought as a superposition of different ideal spectra which smears out the spectrum details and broadens the peaks. Thus, the experimental curves differ significantly from the simulated ones, which assume a perfect substrate with a depth–only dependent deformation. As an example of this problem, in Figure 2b we report a typical comparison between experimental data and a simulated spectrum. The mismatch profile was modeled using 8 lamellae with thicknesses of 50 and 25 nm and mismatch levels determined *assuming a proportionality relation between the defect profile obtained by SRIM simulations and the deformation one. The proportionality constant was adjusted by hand until a reasonable agreement was obtained [da modificare in base ai risultati delle nuove simulazioni di Marco]*. It can be seen that even if some qualitative similarities are present, the two spectra cannot be compared. For this reason it was not possible to perform a fitting of experimental data and reconstruct a reliable deformation profile. In any case, a qualitative comparison of the simulated rocking curves allows us to estimate an order of magnitude for the mismatch induced by the implantation process, which is in the $10^{-4}$–$10^{-3}$ range and has a positive sign (i.e. the lattice expands, as anticipated by the reciprocal lattice map analysis).

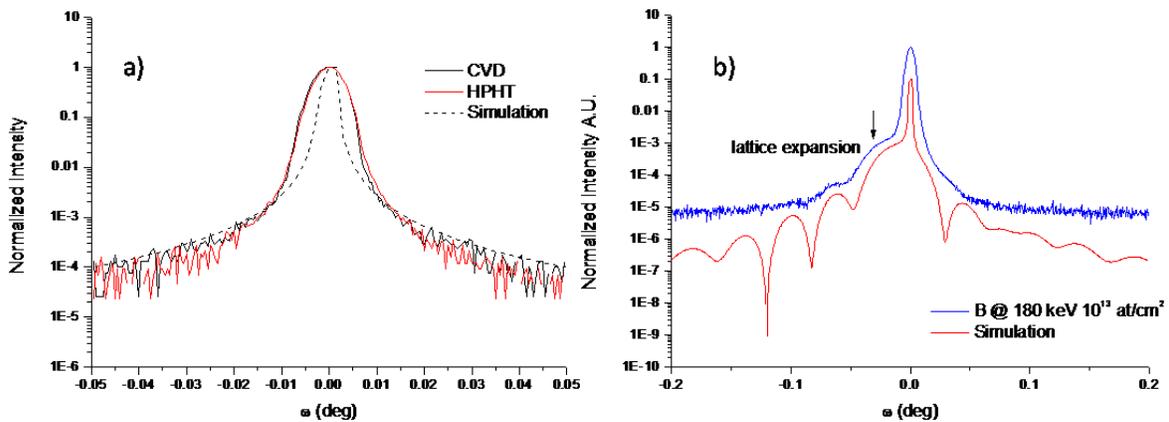

Fig. 2: a) Comparison between rocking curves of the HPHT and CVD substrates and a reference simulation. (0 0 4) lattice planes. b) experimental and simulated rocking curves for the symmetrical 004 reflection for CVD sample implanted with B@180keV, $10^{13}$ ions/cm$^2$.

## 4.2. Surface swelling measurements

Experimental surface swelling data for 1.8 MeV He implantations for fluences varying between $10^{16}$ ions cm$^{-2}$ and $2.5 \cdot 10^{17}$ ions cm$^{-2}$ before and after annealing at 1000 °C are compared to simulation data. Experimental measurements show that after annealing the swelling values decrease and tend to zero for small fluences (below $F=1.4 \cdot 10^{16}$ ions cm$^{-2}$), and the opposite occurs in the large fluence range. This indicates that for fluences below this threshold value, annealing at 1000 °C causes a near complete restoration of the pristine diamond crystal, whilst for fluences above this value graphitization induces a greater swelling. This observation can be exploited to determine $N_C$ from eq.(1), as the vacancy density value corresponding to the threshold fluence value. A value of $N_C \cong 2 \cdot 10^{22}$ vacancies cm$^{-3}$ is therefore estimated for 1.8 MeV He implantations. This value is consistent with previous measurements cited in the literature for different types of implantations (ion types and energies) [*inserire tutte le referenze*], and is somewhat smaller than expected for measurements relative to similar implantations [2], while being closer to the values reported by [*referenza a Uzan-Saguy*]. The reason for this could be the approach described in Section 2, where saturation effects are duly accounted for in the vacancy density for increasing fluence, as expressed by Eq.(1). This effect was not considered in the determination of $N_C$ in previous literature. The derived parameter $N_C$ is therefore used in FEM simulations.

Figure 3 shows experimentally measured and numerically calculated surface swelling profiles for $F=4.18 \cdot 10^{16}$ ions cm$^{-2}$. For this above-threshold fluence value, swelling increases after specimen annealing. Numerical simulations capture this behavior very satisfactorily.

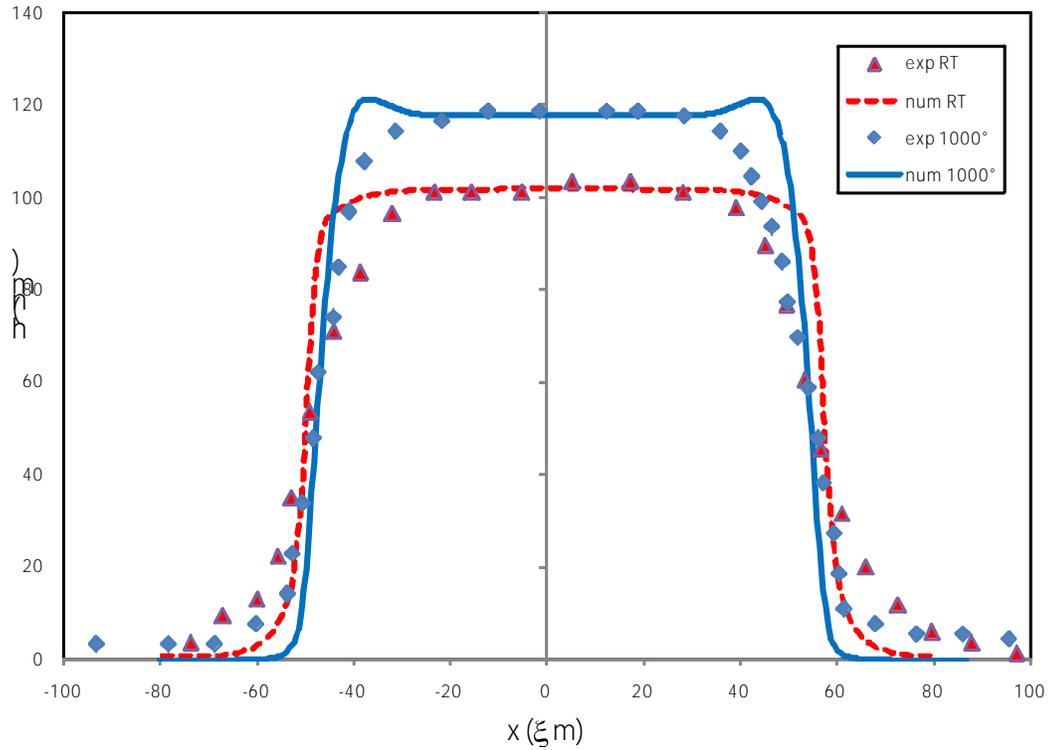

Fig. 3. Experimentally measured ("exp") and numerically computed ("num") surface swelling profiles $h(x)$ before (RT=Room Temperature) and after annealing (1000 °C) for 1.8MeV He ions.

Figure 4 shows the variation of experimental and numerically calculated swelling values as a function of fluence $F$, before and after annealing. Error bars for experimental data are estimated at ±5% on fluence values and ±5nm on displacement values. The best fit of numerical data for samples before annealing is obtained for $\alpha_{1.8\ MeV\ He} = 7\cdot10^{22}$ vac cm$^{-3}$, which is a correction on previously reported values [7], due to the wider fluence range considered for the fit on experimental data. Numerical data fit experimental values very satisfactorily in the low-fluence range, whilst there is some discrepancy for higher fluences. This could be due to the simplified assumptions in deriving Eq.(1), which is based on a linear relationship between the recombination probability for a vacancy in a damage cascade and the vacancy density itself [7]. Further data analysis on a greater number of implantation examples is needed to address this issue.

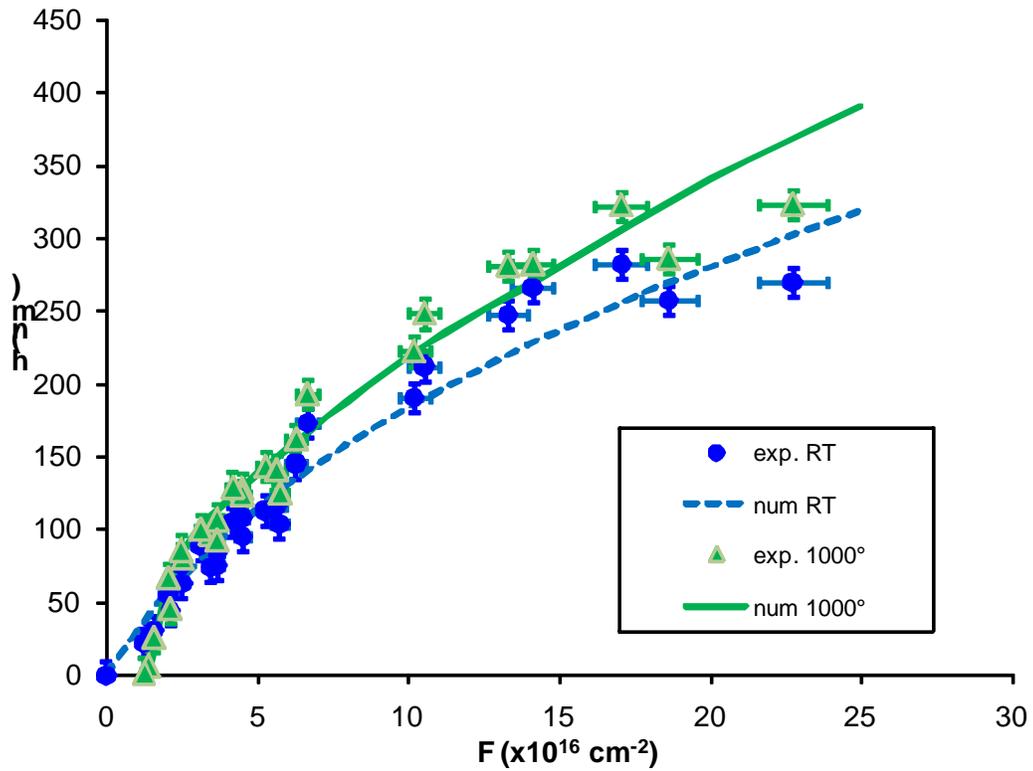

Fig. 4. Experimentally measured ("exp") and numerically computed ("num") surface swelling values $h$ as a function of implantation fluence $F$ before ("RT"=Room Temperature) and after annealing ("1000°") for He@1.8MeV ions.

**Conclusions**

…

**Acknowledgments**

The authors wish to thank Dr. Paolo Schina for specimen preparation at the Olivetti I-JET facilities at Arnad (Aosta). This work is supported by the "Accademia Nazionale dei Lincei – Compagnia di San Paolo" Nanotechnology grant and by "DANTE" and "FARE" experiments of "Istituto Nazionale di Fisica Nucleare" (INFN), which are gratefully acknowledged..

**Figure captions**

Fig. 1. a) Vacancy density vs. specimen depth z for diamond implanted with He@1.8MeV: two examples are shown, above and below the graphitization threshold; b) corresponding mass density before and after annealing at 1000°C.

Fig. 2. a) Comparison between rocking curves of the HPHT and CVD substrates and a reference simulation. (0 0 4) lattice planes. b) experimental and simulated rocking curves for the symmetrical 004 reflection for CVD sample implanted with B@180keV, $10^{13}$ at/cm$^2$.

Fig. 3. Experimentally measured ("exp") and numerically computed ("num") surface swelling profiles $h(x)$ before (RT=Room Temperature) and after annealing (1000°) for He@1.8MeV ions.

Fig. 4. Experimentally measured ("exp") and numerically computed ("num") surface swelling values $h$ as a function of implantation fluence $F$ before ("RT"=Room Temperature) and after annealing ("1000°") for He@1.8MeV ions.